%% file: main_v2.tex
\documentclass{article}

\PassOptionsToPackage{numbers, compress}{natbib}
\usepackage[preprint]{neurips_2023}

\usepackage[utf8]{inputenc} % allow utf-8 input
\usepackage[T1]{fontenc}    % use 8-bit T1 fonts
\usepackage{hyperref}       % hyperlinks
\usepackage{url}            % simple URL typesetting
\usepackage{booktabs}       % professional-quality tables
\usepackage{amsfonts}       % blackboard math symbols
\usepackage{nicefrac}       % compact symbols for 1/2, etc.
\usepackage{microtype}      % microtypography
\usepackage{xcolor}         % colors
\usepackage{amsmath}
\usepackage{amssymb}
\usepackage{mathtools}
\usepackage{amsthm}
\usepackage{multirow}
\usepackage{wrapfig}
\usepackage[capitalize,noabbrev]{cleveref}
\usepackage{graphicx}
\usepackage{subfigure} % for professional tables
\usepackage{float}
\usepackage{algorithm} 
\usepackage{algorithmic}
\usepackage{textcomp} 
\usepackage{listings} 
\usepackage{enumitem}
\usepackage{utfsym}
\usepackage{times}
\usepackage{latexsym}
\usepackage{fancyvrb}
\usepackage{CJKutf8}
\usepackage{makecell}

\newcommand{\commentout}[1]{}

\title{FireRedTTS3: Unified Speech Generation and Editing with Semantically Enriched Speech Representations}

\author{  
Feiyu Shen, Kun Xie, Yichen Wu, Ziqi Dai, Yichen Han, Junjie Li,\\  
\textbf{Xuelong Geng, Fenglong Xie, Lei Xie, Xu Tang, Yao Hu} \\ [6pt]
Xiaohongshu
}

\begin{document}
\maketitle

% Abstract
\begin{abstract}
Recent continuous autoregressive TTS models operate directly on continuous speech representations, preserving rich acoustic details while leveraging the instruction-following capabilities of text LLMs. This paradigm opens new possibilities for voice cloning, instruction-controlled voice design, and speech editing, but remains susceptible to error accumulation during autoregressive generation. 
Existing solutions often require additional semantic modules, multi-stage tokenizer training pipelines, or complex autoregressive architectures.
In this work, we propose FireRedTTS3, a simple yet effective speech generation and editing framework that mitigates error accumulation at the representation level. Specifically, we leverage a frozen Audio Encoder trained on diverse speech understanding tasks as a semantic teacher to regularize the audio feature space. This improves text-speech alignment and stabilizes autoregressive generation while keeping the overall system simple.
FireRedTTS3 provides two variants: FireRedTTS3-Base for multilingual and multi-dialect zero-shot voice cloning, and FireRedTTS3-Instruct for unified voice cloning, instruction-controlled voice design, and speech editing. Experiments show that FireRedTTS3-Base achieves the best average speech intelligibility and speaker similarity among compared systems on Seed-TTS-Eval and MiniMax-MLS-Test, while FireRedTTS3-Instruct outperforms competing systems on InstructTTSEval and Ming-Freeform-Audio-Edit. These results demonstrate that semantically enriched continuous speech representations, combined with a simple architecture, enable stable, controllable, and high-fidelity speech generation and editing. Code and models are available at \url{https://github.com/FireRedTeam/FireRedTTS3}.
\end{abstract}

% Introduction
\section{Introduction}
\label{sec:intro}

Current text-to-speech (TTS) systems have achieved impressive performance in zero-shot voice cloning. However, user demands are increasingly moving beyond cloning toward instruction-controlled voice design and flexible speech editing. These tasks require models to ground natural language instructions in speech properties, modify content or acoustic attributes with precise localization, and faithfully preserve unedited regions. Meeting these requirements depends on speech representations that encode both semantic information and acoustic details, together with generation models capable of following natural language instructions and mapping user intents to concrete operations.

Existing approaches still fall short of these requirements. Flow-matching-based methods~\cite{f5tts,e2tts,voicebox,matchatts,tangoflux,audioldm2,acestep} can generate high-quality audio from textual transcriptions or descriptions. However, they rely on pretrained text encoders for text understanding, and their non-autoregressive architectures limit their ability to leverage text LLMs for instruction-following. 
Another line of work~\cite{cosy1,cosy2,cosy3,indextts,indextts2,indextts25,fireredtts1,fireredtts1s,xcodec,xcodec2,xytokenizer,mossaudiotokenizer,fireredtts2,fishaudio2} discretizes speech into tokens using vector quantization (VQ)~\cite{vq} or residual vector quantization (RVQ)~\cite{rvq}. 
This enables autoregressive modeling and allows speech models to benefit from LLM-based instruction tuning. However, quantization inevitably removes fine-grained acoustic details and can cause distortions in acoustically sensitive tasks such as speech editing.

To retain fine-grained acoustic details while preserving autoregressive modeling, continuous autoregressive methods~\cite{ditar,vibevoice,minguniaudio,mingomnitts,dotstts} bypass quantization and directly model continuous speech representations. By reformulating discrete token prediction as latent denoising, they reuse the autoregressive modeling paradigm and can inherit the instruction-following capabilities of text LLMs; we refer to this formulation as the LLM-DiT framework. This paradigm opens up new possibilities for instruction-controlled speech generation and editing. However, continuous features are defined in an unbounded space, where small prediction errors can accumulate across autoregressive steps and cause severe degradation, such as timbre shifts and prosody collapse~\cite{pasini2024continuous,semavoice}.

Prior work~\cite{vibevoice,minguniaudio,dotstts,voxcpm1,voxcpm2} addresses this issue in two directions: enhancing text-speech alignment with semantically enriched continuous representations, and regularizing the continuous feature space in subsequent autoregressive modeling. 
For the first direction, Ming-UniAudio~\cite{minguniaudio} extracts semantic features from VAE features with an additional module; VibeVoice~\cite{vibevoice} uses a separate semantic VAE to guide LLM modeling; and dots.tts~\cite{dotstts} adds an ASR objective during VAE training for explicit semantic supervision. 
For the second direction, VoxCPM~\cite{voxcpm1,voxcpm2} introduces an FSQ~\cite{fsq} bottleneck to regularize continuous features in autoregressive modeling. Although effective, the former often requires additional semantic modules or multi-stage tokenizer training pipelines, while the latter increases architectural complexity.

In this work, we propose FireRedTTS3, an LLM-DiT speech synthesis framework that mitigates error accumulation at the representation level, without additional semantic modules, multi-stage tokenizer training pipelines, or complex architectures. We develop two variants: FireRedTTS3-Base, which supports zero-shot voice cloning across 24 languages and 21 Chinese dialects; and FireRedTTS3-Instruct, which unifies voice cloning, instruction-controlled voice design, and speech editing within a single model. 

Our main contributions are summarized as follows.
\begin{itemize}[leftmargin=1cm]
\item \textbf{Semantically enriched continuous representations.} We introduce RedAE, a continuous speech tokenizer that incorporates semantic supervision from a frozen Audio Encoder pretrained on diverse speech understanding tasks. This design injects semantic information into the latent space without additional semantic modules or multi-stage tokenizer training. The resulting representations stabilize downstream LLM-DiT modeling, enabling FireRedTTS3 to maintain a simple autoregressive architecture.

\item \textbf{Robust multilingual and multi-dialect voice cloning.} FireRedTTS3-Base enables zero-shot voice cloning across 24 languages and 21 Chinese dialects. It achieves the best average speech intelligibility and speaker similarity among the compared systems on both Seed-TTS-Eval~\cite{seedtts} and MiniMax-MLS-Test~\cite{minimaxspeech} benchmarks.

\item \textbf{Unified instruction-controlled speech generation and editing.}  FireRedTTS3-Instruct unifies zero-shot voice cloning, instruction-controlled voice design, and instruction-controlled speech editing within a single model. It achieves the best performance on InstructTTSEval~\cite{instructttseval} and Ming-Freeform-Audio-Edit, highlighting its potential as a unified framework for instruction-based speech generation and manipulation.
\end{itemize}

\section{Method}
FireRedTTS3 consists of two key components: the RedAE Tokenizer, a semantically enriched speech tokenizer, and a lightweight LLM-DiT generation framework. We instantiate this design into two variants: FireRedTTS3-Base, which targets multilingual and multi-dialect voice cloning, and FireRedTTS3-Instruct, which enables instruction-controlled voice design and speech editing.

\begin{figure}[htp]
\centering
\includegraphics[width=\linewidth]{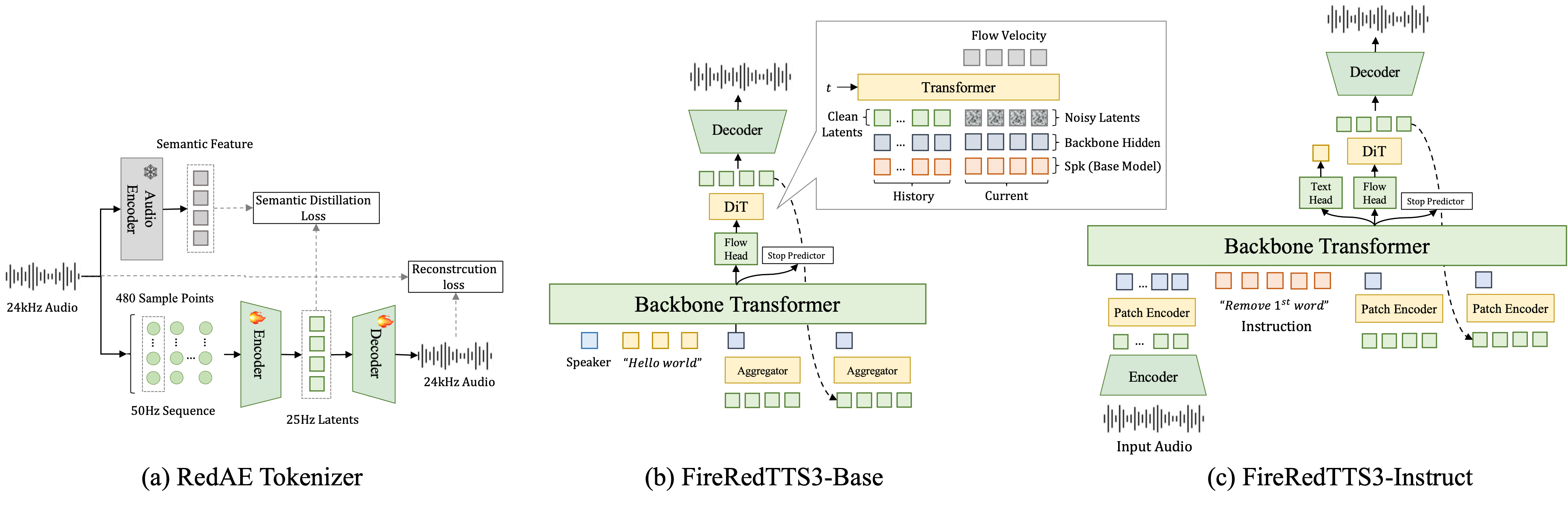}
\caption{An overview of FireRedTTS3, including (a) the RedAE Tokenizer with semantic supervision, (b) FireRedTTS3-Base for multilingual and multi-dialect voice cloning, and (c) FireRedTTS3-Instruct for voice cloning, instruction-controlled voice design, and speech editing.}
\label{img:tts3_arch}
\end{figure}

\subsection{RedAE Tokenizer}
\label{sec:redae}

RedAE provides semantically enriched continuous speech representations for stable LLM-DiT modeling. It learns a latent space that preserves acoustic details for reconstruction, while using semantic features from a pretrained Audio Encoder to improve text alignment.

As shown in Figure~\ref{img:tts3_arch}(a), RedAE adopts a hybrid autoencoder architecture with an Encoder, a Decoder, and a frozen Audio Encoder. The input 24 kHz waveform is segmented into non-overlapping 480-sample frames, yielding a 50 Hz sequence. Two cascaded Qwen3~\cite{qwen3}-style Transformers then process the sequence: the first performs contextual modeling at 50 Hz, and the second downsamples it to 25 Hz through attention-based pooling. Specifically, every two consecutive frames are grouped into a patch with a learnable special token prepended, whose hidden state is used as the 25 Hz RedAE representation.

To preserve reconstruction fidelity, RedAE omits KL regularization~\cite{vae}, avoiding acoustic over-compression, and introduces semantic distillation to stabilize downstream LLM-DiT modeling. We first train FireRedAudio, an audio understanding model, on diverse tasks such as ASR and speaker verification. This enables its Audio Encoder to capture linguistic semantics and speaker-related acoustic cues. During RedAE training, the Audio Encoder is frozen and serves as a semantic teacher; after tokenizer training, it is discarded and not used in downstream LLM-DiT modeling. We minimize the MSE between the 25 Hz RedAE representations and the teacher features, encouraging the latents to be semantically grounded. The Decoder, also a Qwen3-style Transformer, upsamples the 25 Hz latents to 50 Hz and predicts the STFT spectrum, which is converted to a 24 kHz waveform via iSTFT.

With the Audio Encoder frozen as the semantic teacher, RedAE jointly optimizes its Encoder and Decoder in a single training stage, without any additional trainable semantic branch or multi-stage tokenizer pipeline. Specifically, RedAE is trained under a GAN framework, with discriminators following X-Codec~\cite{xcodec}. The objective includes both the discriminator loss $\mathcal{L}_{dis}$ and the generator loss $\mathcal{L}_{gen}$, where

\begin{equation}
\mathcal{L}_{gen} =
\lambda_{adv}\mathcal{L}_{adv}+
\lambda_{mel}\mathcal{L}_{mel}+
\lambda_{fm}\mathcal{L}_{fm}+
\lambda_{sem}\mathcal{L}_{sem}.
\end{equation}
Here, $\mathcal{L}_{adv}$, $\mathcal{L}_{mel}$, $\mathcal{L}_{fm}$, and $\mathcal{L}_{sem}$ denote the adversarial, multi-scale Mel reconstruction, feature matching, and semantic supervision losses, respectively. To improve generalization, RedAE is trained for 550k steps on 32 H800 GPUs using 500k hours of diverse audio, consisting of clean speech (50\%), noisy speech (25\%), sound effects (10\%), and music (15\%).

% ---
\subsection{FireRedTTS3}
Built on RedAE representations, FireRedTTS3 adopts a lightweight LLM-DiT framework. Instead of frame-level autoregression, it models latent patches to reduce sequence length while preserving local context~\cite{ditar}. As shown in Figure~\ref{img:tts3_arch}(b) and (c), FireRedTTS3-Base and FireRedTTS3-Instruct share three components: an Aggregator, a Backbone Transformer, and a DiT module. The Aggregator compresses RedAE representations into latent patches, the Backbone autoregressively models text tokens and latent patches, and the DiT generates RedAE latents conditioned on Backbone hidden states.

The Aggregator is a full-attention Transformer. It compresses 25 Hz RedAE representations into 6.25 Hz patches using the attention-based pooling described in Section~\ref{sec:redae}. The Backbone Transformer is initialized from a pretrained Qwen3~\cite{qwen3} text model to inherit text-understanding capabilities. Its final-layer hidden states condition the DiT module and are also used by a binary classifier for stop prediction.

The DiT module is also a full-attention Transformer, with timesteps injected through AdaLN~\cite{adaln}. It performs patch-level denoising conditioned on Backbone hidden states. At each autoregressive step, its input consists of a noisy 4-frame current patch and 8 frames of clean historical latents, concatenated with their corresponding Backbone conditions. FireRedTTS3-Base additionally uses a speaker embedding to improve speaker consistency. For classifier-free guidance (CFG), we randomly drop the Backbone condition with a probability of 0.1 during training. When speaker conditioning is used, the speaker condition is dropped with the same probability. Clean historical latents are always retained to maintain temporal continuity.

% ---
\textbf{FireRedTTS3-Base} is designed for multilingual and multi-dialect zero-shot voice cloning. Its Backbone Transformer is initialized from Qwen3-1.7B-Base\footnote{\url{https://huggingface.co/Qwen/Qwen3-1.7B-Base}}. To improve speaker similarity, we extract speaker embeddings using the pretrained CAM++\footnote{\url{https://modelscope.cn/models/iic/speech_campplus_sv_en_voxceleb_16k}}~\cite{campp} model. The speaker embedding is prepended to the Backbone input sequence and also serves as a speaker condition for the DiT module. A language tag is prepended to the input text to indicate the target language.

FireRedTTS3-Base is optimized with a flow-matching loss and a stop-prediction loss:
\begin{equation}
\mathcal{L}_{base} = \mathcal{L}_{flow} + \lambda_{stop}\mathcal{L}_{stop}.
\end{equation}
Training is conducted in two stages. The first stage uses 2.6M hours of Chinese and English speech for 170k steps to establish zero-shot voice cloning capability. The second stage continues training on 560k hours of speech covering 24 languages and 21 Chinese dialects, extending the model to multilingual and multi-dialect voice cloning.

% ---
\textbf{FireRedTTS3-Instruct} unifies voice cloning, instruction-controlled voice design, and speech editing. Its Backbone Transformer is initialized from Qwen3-1.7B\footnote{\url{https://huggingface.co/Qwen/Qwen3-1.7B}} to inherit its instruction-following capabilities. We adopt the ChatML format and use task-specific system prompts to distinguish different tasks.

For instruction-controlled voice design and speech editing, the model first converts the input instruction into a structured textual plan, which then guides speech synthesis. Specifically, for voice design, the plan converts free-form instructions into a sequence of 12 concrete acoustic attributes. For speech editing, it expands the instruction into the target transcription and an edit-region mask, enabling localized modifications while preserving unedited regions~\cite{minguniaudio}. We retain the text head of Qwen3-1.7B to support this intermediate planning process. Unlike the Base model, the Instruct model does not use explicit speaker embeddings or language tags.

The Instruct model is optimized with a flow-matching loss, a text loss, and a stop-prediction loss:
\begin{equation}
\mathcal{L}_{inst} =
\mathcal{L}_{flow} +
\mathcal{L}_{text} +
\lambda_{stop}\mathcal{L}_{stop}.
\end{equation}

FireRedTTS3-Instruct follows the same first-stage training procedure as the Base model. In the second stage, it is further trained on 330k hours of voice design and speech editing data for 40k steps, enabling instruction-controlled speech generation and editing.

% Results
\section{Results}

\subsection{Experimental Setup}
We evaluate FireRedTTS3-Base and FireRedTTS3-Instruct on four benchmarks covering multilingual voice cloning, instruction-controlled voice design, and speech editing. For intelligibility, we report word error rate (WER) or character error rate (CER) depending on the language. Speaker similarity is measured with a fine-tuned WavLM-Large~\cite{wavlm} model.

\textbf{Seed-TTS-Eval}\footnote{\url{https://github.com/BytedanceSpeech/seed-tts-eval}} evaluates Chinese and English zero-shot voice cloning, including Test-EN, Test-ZH, and Test-Hard. We use Whisper-large-v3~\cite{whisper} for English WER, Paraformer-ZH~\cite{paraformer} for Chinese CER, and WavLM-Large for speaker similarity.

\textbf{MiniMax-MLS-Test}\footnote{\url{https://huggingface.co/datasets/MiniMaxAI/TTS-Multilingual-Test-Set}} evaluates multilingual voice cloning across 24 languages. We use Paraformer-ZH for Mandarin Chinese and Whisper-large-v3 for the other languages. CER is reported for Chinese, Cantonese, Japanese, Korean, Arabic, Vietnamese, Hindi, Thai, and Greek, while WER is reported for the remaining languages.

\textbf{InstructTTSEval}\footnote{\url{https://huggingface.co/datasets/CaasiHUANG/InstructTTSEval}} evaluates instruction-controlled voice design in Chinese and English, covering three instruction types: acoustic-parameter specification (APS), descriptive-style directive (DSD), and role-play (RP). Since the official evaluation toolkit uses Gemini-2.5-pro-preview, which is inaccessible, we use Gemini-2.5-pro~\cite{gemini25} to score the consistency between generated speech and input instructions for all compared systems.

\textbf{Ming-Freeform-Audio-Edit}\footnote{\url{https://github.com/inclusionAI/Ming-Freeform-Audio-Edit}} evaluates instruction-controlled speech editing, covering both semantic and acoustic editing tasks. Semantic editing includes insertion, deletion, and substitution, evaluated in both the basic setting with template-based instructions and the open setting with free-form instructions. Acoustic editing evaluates the control of speaking rate, pitch, and volume. For semantic editing, we report WER/CER, editing accuracy, and speaker similarity. For acoustic editing, we report WER/CER and speaker similarity, and additionally use relative duration error (RDE) for speaking rate control and relative amplitude error (RAE) for volume control.

\subsection{Zero-shot Voice Cloning on Seed-TTS-Eval}

% Seed-TTS-Eval Table
\begin{table}[htp]
\centering
\caption{Zero-shot voice cloning results on Seed-TTS-Eval. \textbf{Bold} and \underline{underline} denote the best and second-best results, respectively. All results are obtained using the official evaluation scripts.}
\resizebox{0.95\textwidth}{!}{
\input{table/base_seedttseval}
}
\label{tab:base_seedttseval}
\end{table}

As shown in Table~\ref{tab:base_seedttseval}, FireRedTTS3-Base achieves the lowest average error rate and the highest average speaker similarity among the compared systems. For intelligibility, it ranks second on Test-EN and Test-Hard and remains competitive on Test-ZH. For speaker similarity, it obtains the best SIM scores on both Test-ZH and Test-EN, and ranks second on Test-Hard.

These results indicate that the semantically enriched RedAE representations serve as a stable modeling target for the LLM-DiT framework, facilitating robust text-speech alignment. Moreover, because RedAE representations avoid the acoustic information loss introduced by quantization, the model can preserve more acoustic details, leading to superior voice cloning similarity.

\subsection{Multilingual Zero-shot Voice Cloning}

\begin{table}[htp]
\centering
\setlength{\tabcolsep}{4pt}
\caption{Per-language intelligibility and speaker similarity on MiniMax-MLS-Test reported in \%. \textbf{Bold} and \underline{underline} indicate the best and second-best results, respectively. The high Cantonese CER is attributed to the limited recognition capability of Whisper-large-v3.}
\resizebox{\textwidth}{!}{
\input{table/base_minimax}
}
\label{tab:base_minimax}
\end{table}

As shown in Table~\ref{tab:base_minimax}, FireRedTTS3-Base achieves the lowest average error rate across all 24 languages and ranks first or second in 8 languages. The high Cantonese error rates observed across all systems are mainly due to the limited Cantonese recognition capability of Whisper-large-v3, and therefore may not faithfully reflect synthesis quality. Notably, Portuguese and Ukrainian are not covered in the training data of either the Audio Encoder or RedAE. Nevertheless, FireRedTTS3-Base achieves competitive error rates on both languages, demonstrating the generalization capability of RedAE representations to unseen languages.

In terms of speaker similarity, FireRedTTS3-Base ranks first or second in 22 out of 24 languages and achieves the highest average similarity score. These results further confirm the advantage of RedAE representations in preserving acoustic details across different languages.

\subsection{Instruction-Controlled Voice Design}

% InstructTTSEval Table
\begin{table}[htp]
\centering
\caption{Instruction-following accuracy on InstructTTSEval. Since Gemini-2.5-pro-preview used by the official evaluation toolkit is inaccessible, all results are evaluated with Gemini-2.5-pro. APS, DSD, and RP denote acoustic-parameter specification, descriptive-style directive, and role-play, respectively. \textbf{Bold} denotes the best result.}
\resizebox{0.9\textwidth}{!}{
\input{table/instruct_voicedesign}
}
\label{tab:instruct_voicedesign}
\end{table}

As shown in Table~\ref{tab:instruct_voicedesign}, FireRedTTS3-Instruct achieves the best results on both Chinese and English across APS, DSD, and RP tasks. For APS, the intermediate textual planning step extracts explicit acoustic attributes from complex parameter-style instructions. For DSD and RP, it translates abstract style descriptions or role specifications into concrete acoustic attributes. This reduces the ambiguity of natural language instructions before speech generation. Beyond instruction parsing, FireRedTTS3 further learns a stable mapping from textual acoustic descriptions to acoustic attributes in RedAE representations, enabling effective control over specific acoustic dimensions.

\subsection{Instruction-Controlled Speech Editing}

% Ming-Freeform-Audio-Edit Acoustic Table
\begin{table}[htp]
\centering
\caption{Instruction-controlled acoustic editing results on Ming-Freeform-Audio-Edit. \textbf{Bold} denotes the best result.}
\resizebox{0.8\textwidth}{!}{
\input{table/instruct_aco_edit}
}
\label{tab:instruct_aco_edit}
\end{table}

For acoustic editing, as shown in Table~\ref{tab:instruct_aco_edit}, FireRedTTS3 demonstrates effective control over speaking rate, pitch, and volume. It achieves the desired acoustic modifications while maintaining high speech fidelity.

As shown in Table~\ref{tab:instruct_sem_edit}, for semantic editing, FireRedTTS3 performs consistently on deletion, insertion, and substitution tasks, and generalizes well to open-ended scenarios with free-form instructions. Its lower overall and unedited-region WERs, higher editing accuracy, and stable speaker similarity indicate that it can accurately perform target edits while preserving the remaining content and speaker characteristics. 

% Ming-Freeform-Audio-Edit Semantic Table
\begin{table}[htp]
\centering
\caption{Instruction-controlled semantic editing results on Ming-Freeform-Audio-Edit. \textbf{Bold} denotes the best result.}
\resizebox{0.9\textwidth}{!}{
\input{table/instruct_sem_edit}
}
\label{tab:instruct_sem_edit}
\end{table}

These results suggest that RedAE representations combine strong semantic modeling with rich acoustic detail preservation. They allow a simple LLM-DiT framework to achieve effective feature alignment and accurately localize both semantic and acoustic editing targets. Meanwhile, they help preserve unedited regions with high fidelity.

% Conclusion
\section{Conclusions}
In this work, we propose FireRedTTS3, which mitigates error accumulation in continuous autoregressive modeling through semantically enriched speech representations. We achieve this by introducing RedAE, a continuous speech tokenizer that injects semantic information from a frozen Audio Encoder pretrained on diverse speech understanding tasks. This semantic supervision improves text-speech alignment and stabilizes subsequent autoregressive modeling. Furthermore, RedAE keeps tokenizer training simple: its Encoder and Decoder are jointly optimized in a single stage, without additional semantic modules or multi-stage training pipelines. Built on the RedAE representations, FireRedTTS3 further adopts a simple LLM-DiT architecture and provides two variants. FireRedTTS3-Base supports multilingual and multi-dialect zero-shot voice cloning, while FireRedTTS3-Instruct extends the same framework to instruction-controlled voice design, as well as semantic and acoustic speech editing. Experimental results show that FireRedTTS3 achieves strong performance across all evaluated tasks, demonstrating the effectiveness of semantically enriched speech representations for stable and controllable speech generation and editing.

% \newpage
\bibliographystyle{unsrt}
\bibliography{refs}

\end{document}

%% file: table/base_seedttseval.tex
\begin{tabular}{lccccccccc}
\toprule
\multirow{2}{*}{\textbf{Model}}
  & \multicolumn{2}{c}{\textbf{Test-EN}}
  & \multicolumn{2}{c}{\textbf{Test-ZH}}
  & \multicolumn{2}{c}{\textbf{Test-Hard}}
  & \multicolumn{2}{c}{\textbf{Average}} \\
\cmidrule(lr){2-3} \cmidrule(lr){4-5} \cmidrule(lr){6-7} \cmidrule(lr){8-9}
  & WER(\%)$\downarrow$ & SIM(\%)$\uparrow$
  & CER(\%)$\downarrow$ & SIM(\%)$\uparrow$
  & CER(\%)$\downarrow$ & SIM(\%)$\uparrow$
  & WER/CER(\%)$\downarrow$ & SIM(\%)$\uparrow$ \\
\midrule
CosyVoice3-1.5B & 2.22 & 72.0 & 1.12 & 78.1 & \textbf{5.83} & 75.8 & \underline{3.06} & 75.3 \\
DiTAR           & 1.69 & 73.5 & 1.02 & 75.3 & -- & -- & -- & -- \\
F5-TTS          & 2.00 & 67.0 & 1.53 & 76.0 & 8.67 & 71.3 & 4.10 & 71.4 \\
FireRedTTS2     & 1.95 & 66.5 & 1.14 & 73.6 & 8.98 & 70.3 & 4.02 & 70.1 \\
IndexTTS2       & 2.23 & 70.6 & 1.03 & 76.5 & 7.12 & 75.5 & 3.46 & 74.2 \\
MegaTTS3\cite{megatts3}        & 2.79 & \underline{77.1} & 1.52 & 79.0 & -- & -- & -- & -- \\
MiniMax-Speech  & 1.65 & 69.2 & \textbf{0.83} & 78.3 & -- & -- & -- & -- \\
Qwen3-TTS\cite{qwen3tts}       & \textbf{1.23} & 71.7 & 1.22 & 77.0 & 6.76 & 74.8 & 3.07 & 74.5 \\
Seed-TTS        & 2.25 & 76.2 & 1.12 & 79.6 & 7.59 & 77.6 & 3.65 & 77.8 \\
VibeVoice       & 3.04 & 68.9 & 1.16 & 74.4 & -- & -- & -- & -- \\
VoxCPM2         & 1.84 & 75.3 & \underline{0.97} & 79.5 & 8.13 & 75.3 & 3.65 & 76.7 \\
dots.tts(Pre.)*        & 1.80 & 77.0 & \underline{0.97} & \underline{80.4} & 6.65 & \textbf{78.8} & 3.14 & \underline{78.7} \\
\midrule
FireRedTTS3-Base     & \underline{1.64} & \textbf{77.2} & 1.01 & \textbf{80.9} & \underline{6.50} & \underline{78.4} & \textbf{3.04} & \textbf{78.8} \\
% \textbf{FireTTS3-Instruct} &               &      &      &               & --    & --   & --   & -- \\
\bottomrule
\multicolumn{9}{l}{\footnotesize * (Pre.) denotes the pretraining checkpoint.} \\
\end{tabular}

%% file: table/base_minimax.tex
\begin{tabular}{l cccccc | cccccc}
\toprule
& \multicolumn{6}{c|}{\textbf{CER/WER(\%)$\downarrow$}} & \multicolumn{6}{c}{\textbf{Speaker Similarity(\%)$\uparrow$}} \\
\cmidrule(lr){2-7} \cmidrule(lr){8-13}
\textbf{Language} & MiniMax & ElevenLabs & VoxCPM2 & FishAudioS2 & dots.tts(Pre.) & FireRedTTS3 & MiniMax & ElevenLabs & VoxCPM2 & FishAudioS2 & dots.tts(Pre.) & FireRedTTS3 \\
\midrule
Arabic     & \textbf{1.67}  & \textbf{1.67} & 13.05 & 3.50          & 37.91 & \underline{1.75}  & 73.6 & 70.6 & \textbf{79.1} & 75.0 & 77.5          & \underline{78.9} \\
Cantonese  & \underline{34.11} & 51.51      & 38.58 & \textbf{30.67}& 37.91 & 40.32             & 77.8 & 67.0 & 83.5          & 80.5 & \textbf{84.7} & \underline{83.9} \\
Chinese    & 2.25  & 16.03 & 1.14  & \textbf{0.73} & 1.08  & \underline{0.91}  & 78.0 & 67.7 & \underline{82.5} & 81.6 & 82.3         & \textbf{84.2} \\
Czech      & 3.88  & \textbf{2.11} & 24.13 & \underline{2.84}         & 5.05  & 3.17  & 79.6 & 68.5 & 78.3          & 79.8 & \underline{83.8} & \textbf{86.1} \\
Dutch      & 1.14  & \textbf{0.80} & \underline{0.91} & 0.99 & 1.20 & 1.15             & 73.8 & 68.0 & 80.8          & 73.0 & \underline{81.4} & \textbf{84.3} \\
English    & 2.16  & 2.34  & 2.29  & \underline{1.62}          & \textbf{1.06} & 2.12  & 75.6 & 61.3 & 85.4          & 79.7 & \textbf{86.9} & \underline{86.8} \\
Finnish    & 4.67  & \underline{2.96}  & \textbf{2.63} & 3.33         & 3.44  & 3.10  & 83.5 & 75.9 & \underline{89.0} & 81.9 & 88.0         & \textbf{89.9} \\
French     & 4.10  & 5.22  & 4.53  & \textbf{3.05} & \underline{3.82} & 5.28           & 62.8 & 53.5 & 73.5          & 69.8 & \underline{78.2} & \textbf{81.0} \\
German     & 1.91  & \underline{0.57} & 0.68 & \textbf{0.55} & 1.03 & 0.69            & 73.3 & 61.4 & \underline{80.3} & 76.7 & 79.5         & \textbf{83.3} \\
Greek      & 2.02  & \textbf{0.99} & 2.84 & 5.74          & 2.97  & \underline{1.24}  & 82.6 & 73.3 & 86.0          & 79.5 & \underline{87.6} & \textbf{89.3} \\
Hindi      & \underline{6.96} & \underline{5.83} & 19.70 & 14.64      & 14.32 & 7.02             & 81.8 & 73.0 & \underline{85.6} & 82.1 & 84.5         & \textbf{87.2} \\
Indonesian & 1.24  & \textbf{1.06} & \underline{1.08} & 1.46 & 2.71 & 1.42             & 72.9 & 66.0 & 80.0          & 76.3 & \underline{80.8} & \textbf{83.3} \\
Italian    & \underline{1.54} & 1.74 & 1.56  & \textbf{1.27} & 3.16 & 2.28             & 69.9 & 57.9 & 78.0          & 74.7 & \textbf{84.5} & \underline{83.6} \\
Japanese   & \underline{3.52}  & 10.65 & 4.63  & \textbf{2.76} & 7.16  & 3.60  & 77.6 & 73.8 & \underline{82.8} & 79.6 & \textbf{83.1}         & \underline{82.8} \\
Korean     & \underline{1.75} & 1.87 & 1.96  & \textbf{1.18} & 5.30 & 2.42             & 77.6 & 70.0 & 83.3          & 81.7 & \underline{84.3} & \textbf{86.6} \\
Polish     & 1.42  & \textbf{0.77} & \underline{1.14} & 1.26          & 2.72  & 1.22  & 80.2 & 72.9 & 88.4          & 81.9 & 87.3          & \textbf{89.8} \\
Portuguese & 1.88  & \underline{1.33} & 1.94 & \textbf{1.14} & 1.64 & 1.79             & 80.5 & 71.1 & \underline{83.7}          & 78.1 & 83.1          & \textbf{86.3} \\
Romanian   & 2.88  & \textbf{1.35} & 21.58 & 10.74        & 3.36  & \underline{1.93}  & \underline{80.9} & 69.9 & 79.7          & 73.3 & \textbf{86.2} & \textbf{86.2} \\
Russian    & 4.28  & 3.88  & 3.63  & \textbf{2.40} & 3.64  & \underline{3.28}  & 76.1 & 67.6 & 81.1          & 79.0 & \underline{83.0} & \textbf{84.7} \\
Spanish    & 1.03  & 1.08  & 1.44  & \textbf{0.91} & \underline{0.96} & 1.21           & 76.2 & 61.5 & 83.1          & 77.6 & \underline{83.9}          & \textbf{86.3} \\
Thai       & \underline{2.70} & 73.94 & 2.96 & 4.23         & 7.45  & \textbf{1.87}   & 80.0 & 58.8 & \textbf{84.0}          & 78.6 & \underline{83.8} & 83.3 \\
Turkish    & 1.52  & \textbf{0.70} & \underline{0.82} & 0.87 & 5.45 & 0.92            & 77.9 & 59.6 & \underline{87.1} & 83.5 & \textbf{87.4} & 86.6 \\
Ukrainian  & 1.08  & \underline{1.00} & 6.32 & 2.30         & 1.61  & \textbf{0.55}   & 73.0 & 64.7 & \underline{79.8} & 74.7 & \textbf{80.5}         & \underline{79.8} \\
Vietnamese & \underline{0.88} & 73.42 & 3.31 & 7.41          & 3.85  & \textbf{0.86}  & 74.3 & 36.9 & 80.6          & 74.0 & \underline{80.7} & \textbf{81.3} \\
\midrule
Average    & \underline{3.77} & 10.95 & 6.79 & 4.40         & 6.60  & \textbf{3.75}   & 76.6 & 65.5 & 82.3          & 78.0 & \underline{83.5} & \textbf{84.8} \\
\bottomrule
\end{tabular}

%% file: table/instruct_voicedesign.tex
\begin{tabular}{lcccccc}
\toprule
\multirow{2}{*}{\textbf{Model}}
  & \multicolumn{3}{c}{\textbf{InstructTTSEval-ZH}}
  & \multicolumn{3}{c}{\textbf{InstructTTSEval-EN}} \\
\cmidrule(lr){2-4} \cmidrule(lr){5-7}
  & APS(\%)$\uparrow$ & DSD(\%)$\uparrow$ & RP(\%)$\uparrow$ & APS(\%)$\uparrow$ & DSD(\%)$\uparrow$ & RP(\%)$\uparrow$ \\
\midrule
MOSS-VoiceGenerator\cite{mossvoicegen}  & 71.6 & 72.5 & 61.3 & 58.8 & 71.8 & 61.6 \\
VoiceSculptor-VD\cite{voicesculptorvoicedesigned}     & 74.6 & 63.5 & 62.0 & --   & --   & --   \\
Ming-Omni-TTS-16B-A3B    & 84.6 & 70.7 & 56.0 & --   & --   & --   \\
Qwen3-TTS-VD\cite{qwen3tts}         & 83.7 & 81.7 & 65.8 & 76.4 & 81.4 & 64.2 \\
\midrule
FireRedTTS3-Instruct & \textbf{85.8} & \textbf{82.0} & \textbf{69.7} & \textbf{80.7} & \textbf{82.3} & \textbf{72.0} \\
\bottomrule
\end{tabular}

%% file: table/instruct_aco_edit.tex
\begin{tabular}{llcc}
\toprule
\begin{tabular}[c]{@{}l@{}}\textbf{Task}\end{tabular}
& \begin{tabular}[c]{@{}l@{}}\textbf{Metric}\end{tabular}
& \begin{tabular}[c]{@{}c@{}}\textbf{Ming-UniAudio-Edit}\\\textbf{ZH | EN}\end{tabular}
& \begin{tabular}[c]{@{}c@{}}\textbf{FireRedTTS3-Instruct}\\\textbf{ZH | EN}\end{tabular} \\
\midrule
\multirow{3}{*}{Speed Alteration}
& WER(\%)$\downarrow$ & 5.88~|~17.53 & \textbf{2.27}~|~\textbf{4.75} \\
& SIM$\uparrow$ & 0.66~|~0.57 & \textbf{0.80}~|~\textbf{0.71} \\
& RDE(\%)$\downarrow$ & 6.36~|~5.92 & \textbf{4.35}~|~\textbf{4.29} \\
\midrule
\multirow{2}{*}{Pitch Alteration}
& WER(\%)$\downarrow$ & 7.45~|~13.37 & \textbf{2.34}~|~\textbf{2.94} \\
& SIM$\uparrow$ & 0.36~|~0.24 & \textbf{0.51}~|~\textbf{0.44} \\
\midrule
\multirow{3}{*}{Volume Alteration}
& WER(\%)$\downarrow$ & 1.71~|~1.35 & \textbf{1.69}~|~\textbf{1.26} \\
& SIM$\uparrow$ & 0.86~|~0.80 & \textbf{0.92}~|~\textbf{0.90} \\
& RAE(\%)$\downarrow$ & 14.9~|~11.7 & \textbf{3.58}~|~\textbf{4.44} \\
\bottomrule
\end{tabular}

%% file: table/instruct_sem_edit.tex
\begin{tabular}{lllcc}
\toprule
\textbf{Task} & \textbf{Setting} & \textbf{Metric}  & \makecell{\textbf{Ming-UniAudio-Edit} \\ \textbf{ZH | EN}}  & \makecell{\textbf{FireRedTTS3-Instruct} \\ \textbf{ZH | EN}} \\
\midrule
\multirow{8}{*}{Deletion} & \multirow{4}{*}{basic} & WER(\%) $\downarrow$ & 11.89 | 14.85 & \textbf{10.51} | \textbf{14.46} \\
 & & SIM$\uparrow$ & \textbf{0.78} | 0.76 & \textbf{0.78} | \textbf{0.79} \\
 & & ACC(\%)$\uparrow$ & \textbf{100.00} | 82.22 & \textbf{100.00} | \textbf{97.78} \\
 & & no-edit WER(\%)$\downarrow$ & 11.49 | 24.26 & \textbf{10.30} | \textbf{23.97} \\
\cmidrule{2-5}
 & \multirow{4}{*}{open} & WER(\%)$\downarrow$ & 22.92 | 27.60 & \textbf{16.31} | \textbf{18.62} \\
 & & SIM$\uparrow$ & \textbf{0.81} | 0.74 & \textbf{0.81} | \textbf{0.78} \\
 & & ACC(\%) $\uparrow$ & 82.92 | 85.00 & \textbf{89.32} | \textbf{89.50} \\
 & & no-edit WER(\%)$\downarrow$ & 17.50 | 35.21 & \textbf{11.69} | \textbf{27.08} \\
\midrule
\multirow{8}{*}{Insertion} & \multirow{4}{*}{basic} & WER(\%)$\downarrow$ & \textbf{3.42} | \textbf{6.63} & 3.62 | 6.84 \\
 & & SIM$\uparrow$ & \textbf{0.83} | 0.79 & \textbf{0.83} | \textbf{0.83} \\
 & & ACC(\%)$\uparrow$ & 80.00 | 71.43 & \textbf{81.18} | \textbf{76.40} \\
 & & no-edit WER(\%)$\downarrow$ & \textbf{3.52} | \textbf{17.70} & 3.80 | 18.23 \\
\cmidrule{2-5}
 & \multirow{4}{*}{open} & WER(\%)$\downarrow$ & \textbf{3.89} | \textbf{7.59} & 4.79 | 9.05 \\
 & & SIM$\uparrow$ & 0.83 | 0.79 & \textbf{0.84} | \textbf{0.83} \\
 & & ACC(\%)$\uparrow$ & \textbf{79.31} | 62.31 & \textbf{79.31} | \textbf{65.83} \\
 & & no-edit WER(\%)$\downarrow$ & \textbf{4.10} | \textbf{18.84} & 5.22 | 20.22 \\
\midrule
\multirow{8}{*}{Substitution} & \multirow{4}{*}{basic} & WER(\%)$\downarrow$ & 4.52 | 8.99 & \textbf{2.92} | \textbf{5.63} \\
 & & SIM$\uparrow$ & 0.82 | 0.78 & \textbf{0.83} | \textbf{0.80} \\
 & & ACC(\%)$\uparrow$ & 78.62 | 59.78 & \textbf{87.42} | \textbf{75.42} \\
 & & no-edit WER(\%)$\downarrow$ & 4.63 | 19.28 & \textbf{3.19} | \textbf{17.05} \\
\cmidrule{2-5}
 & \multirow{4}{*}{open} & WER (\%) $\downarrow$ & 4.56 | 7.64 & \textbf{3.52} | \textbf{6.54} \\
 & & SIM$\uparrow$ & \textbf{0.83} | 0.77 & \textbf{0.83} | \textbf{0.80} \\
 & & ACC(\%)$\uparrow$ & 76.62 | 65.62 & \textbf{86.15} | \textbf{71.48} \\
 & & no-edit WER(\%)$\downarrow$ & 4.75 | \textbf{18.39} & \textbf{3.85} | 18.42 \\
\midrule
\multirow{4}{*}{Average} & \multirow{4}{*}{basic+open} & WER(\%)$\downarrow$ & 8.53 | 12.22 & \textbf{6.97} | \textbf{10.22} \\
 & & SIM$\uparrow$ & \textbf{0.82} | 0.77 & \textbf{0.82} | \textbf{0.80} \\
 & & ACC(\%)$\uparrow$ & 82.91 | 71.06 & \textbf{87.27} | \textbf{78.91} \\
 & & no-edit WER(\%)$\downarrow$ & 7.67 | 22.28 & \textbf{6.49} | \textbf{20.90} \\
\bottomrule
\end{tabular}